\documentclass[aps,amsmath,amssymb,twocolumn,showpacs,nofootinbib]{revtex4}
\usepackage{graphicx}
\usepackage{amssymb}
\newcommand{\mb}[1]{\mbox{\scriptsize #1}}

\begin{document}
\title{Chromomagnetic instability in two-flavor quark matter 
at nonzero temperature}
\author{O. Kiriyama}
\email{kiriyama@th.physik.uni-frankfurt.de}
\affiliation{Institut f\"ur Theoretische Physik, 
J.W.\ Goethe-Universit\"at, D-60438 Frankfurt am Main, Germany}

\begin{abstract}
We calculate the effective potential of the 2SC/g2SC phases 
including vector condensates 
($\langle gA_z^6 \rangle$ and $\langle gA_z^8 \rangle$) and 
study the gluonic phase 
and the single plane-wave Larkin-Ovchinnikov-Fulde-Ferrell state 
at nonzero temperature. 
Our analysis is performed within the framework 
of the gauged Nambu--Jona-Lasinio model. 
We compute potential curvatures with respect to the vector 
condensates and investigate the temperature dependence 
of the Meissner masses squared of gluons of color 4--7 and 8 
in the neutral 2SC/g2SC phases. The phase diagram is presented 
in the plane of temperature and coupling strength. 
The unstable regions for gluons 4--7 and 8 are 
mapped out on the phase diagram. 
We find that, apart from the case of strong coupling, 
the 2SC/g2SC phases at low temperatures 
are unstable against the vector condensation 
until the temperature reaches tens of MeV.
\end{abstract}
\date{\today}
\maketitle

\section{Introduction}
Sufficiently cold and dense quark matter has a rich phase structure; 
during the last decade, significant advances have been 
made in our understandings of color superconductivity \cite{CSC}. 
At asymptotically large quark density, 
studies using the perturbative one-gluon exchange interaction, 
directly based on first principles of QCD, are reliable 
and have clarified the nature of the pairing dynamics of quarks. 
However, if the color-superconducting phase is realized in nature, 
it appears in the interior of compact stars. 
The density regime of interest is, therefore, 
up to a few times nuclear density. 
The investigation of the ground state of quark matter 
in this moderate density regime under conditions relevant 
for the bulk of compact stars (i.e., color and electric charge neutrality 
and $\beta$-equilibrium) has recently been 
of great interest \cite{phased1,phased2,phased3}. 

At the present time, it is known that the electric neutrality condition plays 
a crucial role in the Cooper pairing dynamics in quark matter. 
It enforces a substantial Fermi momentum mismatch on quarks. 
As a consequence, the ordinary BCS state is not always energetically favored 
over other unconventional states with the possibility of 
a crystalline color superconductivity \cite{ABR,Rajagopal2006} 
and a gapless color superconductivity 
(e.g., the gapless 2SC (g2SC) phase \cite{Shovkovy2003} 
and gapless color-flavor-locked (gCFL) phase \cite{Alford2003}). 
However, it turned out that the 2SC/g2SC phases suffers 
from a chromomagnetic instability, indicated by 
a negative Meissner masses squared of some gluons \cite{Huang2004}. 
In the two-flavor case (and at zero temperature), 
gluons of adjoint color 8 have a tachyonic Meissner mass 
in the region where the gap $\Delta$ is less than the chemical 
potential mismatch $\delta\mu$, $\Delta/\delta\mu < 1$. 
On the other hand, the more severe instability 
related to gluons 4--7 emerges 
in the region $\Delta/\delta\mu < \sqrt{2}$ 
(i.e., not only in the g2SC phase but also in the 2SC phase). 
The chromomagnetic instability was also found 
in the gCFL phase \cite{Casalbuoni2004,Alford2005,Fukushima2005}, 
though the manifestation of the imaginary Meissner masses 
is different from the two-flavor case. 
The question which phase is realized in neutral quark matter 
is therefore still an open question. 

Resolving the chromomagnetic instability and clarifying 
the nature of ground state are pressing issues in the study 
of color superconductors 
\cite{Giannakis2004,Huang2005,Hong2005,Gorbar2005,Gorbar2005b,
Iida2006,Fukush2006,GHMS2006,Hashimoto2006,KRS2006,Giannakis2006}. 
So far, a single plane-wave Larkin-Ovchinnikov-Fulde-Ferrell (LOFF) state 
\cite{LOFF,Giannakis2004} and a gluonic phase \cite{Gorbar2005} 
have been proposed as candidates 
for the solution to the instability in two-flavor color superconductor. 
It is interesting to note that both phases are equivalent to 
the conventional 2SC/g2SC phases with vector condensates 
[i.e., $\langle\vec{A}_8\rangle \neq 0$ for the single plane-wave LOFF 
state and 
$\langle A_z^6 \rangle,~\langle A_0^3 \rangle,~\langle A_z^1 \rangle \neq 0$ 
for the gluonic phase (without loss of generality)]. 
The neutral single plane-wave LOFF state is indeed 
free from the instability 
in the weak-coupling regime \cite{Giannakis2004,Gorbar2005b}, 
but still suffers from the instability 
related to gluons 4--7 in the intermediate- 
and strong-coupling regimes \cite{Gorbar2005b}. 
In addition, it has been indicated that 
a LOFF state with many plane waves is energetically more favored 
than the single plane-wave LOFF state \cite{ABR}. 
(Note that it has recently been demonstrated that, 
in the three-flavor case, a LOFF state 
with realistic crystal structures has much lower free energy 
than the single plane-wave LOFF state \cite{Rajagopal2006}.) 
On the other hand, the gluonic phase could resolve the instability 
associated with gluons 4--7 in the intermediate- and 
the strong-coupling regimes \cite{Gorbar2005,KRS2006}. 
However, its dynamics has been clarified only around the critical point. 
It should be mentioned that a mixed phase \cite{RedRup} 
and, in the three-flavor case, phases 
with spontaneously induced meson supercurrents \cite{supercurrent} 
are also candidates for the solution to the chromomagnetic instability.

Most of the studies of the chromomagnetic instability 
have been restricted to zero temperature. 
However, in order to look at its relevance for the phase diagram, 
one has to investigate the nature of the instability 
at nonzero temperature. 
To our knowledge the instability is weakened 
by thermal effects and should finally vanish 
at sufficiently high temperature \cite{Alford2005,
Fukushima2005,HJZ,Kiriyama2006}. 

In this work, we calculate the effective potential 
of the 2SC/g2SC phases including the vector condensates 
$\langle gA_z^6 \rangle$ and $\langle gA_z^8 \rangle$. 
The Meissner masses squared in the 2SC/g2SC phases are computed 
from the curvature of the effective potential 
with respect to the vector condensates. 
The potential curvature in the direction of 
$\langle gA_z^6 \rangle$ and $\langle gA_z^8 \rangle$ corresponds 
to the Meissner mass squared of gluons of color 4--7 and 8, respectively. 
We also present the phase diagram in the plane of temperature and 
diquark coupling strength and map out the unstable regions 
for gluons 4--7 and 8 on the phase diagram. 
The results presented in this paper are partly overlapping 
with the previous paper \cite{Kiriyama2006}, 
but we shall take a closer look at the chromomagnetic instability 
by means of a detailed investigation of the effective potential. 

This paper is organized as follows. 
In Sec. II, we introduce the effective potential 
of a gauged Nambu--Jona-Lasinio model with vector condensates 
at nonzero temperature. 
We then derive the formulae for the Meissner masses squared 
at nonzero temperature. We present numerical results for gluons 4--7 
and the 8th gluon, separately. 
We also present the phase diagram and map out 
the unstable regions on the diagram. 
Section III is devoted to conclusions.

\section{Meissner masses at nonzero temperature}
\subsection{Gauged Nambu--Jona-Lasinio model with vector condensates}
In order to study the Meissner screening mass of gluons, 
we use a gauged Nambu--Jona-Lasinio (NJL) model 
with massless up and down quarks:
\begin{eqnarray}
{\cal L}&=&\bar{\psi}(iD\hspace{-7pt}/+\hat{\mu}\gamma^0)\psi
+G_D\left(\bar{\psi}i\gamma_5\varepsilon\epsilon^bC\bar{\psi}^T\right)
\left(\psi^T Ci\gamma_5\varepsilon\epsilon^b\psi\right)\nonumber\\
&&-\frac{1}{4}F_{\mu\nu}^{a}F^{a\mu\nu},
\end{eqnarray}
where the quark field $\psi$ carries flavor ($i,j=1,\ldots N_f$ 
with $N_f=2$) and color ($\alpha,\beta=1,\ldots N_c$ with $N_c=3$) 
indices, $C$ is the charge conjugation matrix; 
$(\varepsilon)^{ik}=\varepsilon^{ik}$ and 
$(\epsilon^b)^{\alpha\beta}=\epsilon^{b\alpha\beta}$ 
are the antisymmetric tensors in flavor and color spaces, 
respectively. The covariant derivative and the field-strength tensor 
are defined as
\begin{subequations}
\begin{eqnarray}
D_{\mu} &=& \partial_{\mu}-igA_{\mu}^{a}T^{a},\\
F_{\mu\nu}^{a} &=& \partial_{\mu}A_{\nu}^{a}-\partial_{\nu}A_{\mu}^{a}
+gf^{abc}A_{\mu}^{b}A_{\nu}^{c}.
\end{eqnarray}
\end{subequations}
In order to evaluate loop diagrams we use a three-momentum cutoff 
$\Lambda=653.3$ MeV. This specific choice does not affect 
the qualitative features of the present analysis. 

In $\beta$-equilibrated neutral quark matter, 
the elements of the diagonal matrix of 
quark chemical potentials $\hat{\mu}$ are given by
\begin{subequations}
\begin{eqnarray}
\mu_{ur}&=&\mu_{ug}=\bar{\mu}-\delta\mu,\\
\mu_{dr}&=&\mu_{dg}=\bar{\mu}+\delta\mu,\\
\mu_{ub}&=&\bar{\mu}-\delta\mu-\mu_8,\\
\mu_{db}&=&\bar{\mu}+\delta\mu-\mu_8,
\end{eqnarray}
\end{subequations}
with
\begin{eqnarray}
\bar{\mu}=\mu-\frac{\delta\mu}{3}+\frac{\mu_8}{3},
\quad\delta\mu=\frac{\mu_e}{2}.
\end{eqnarray}
In a gauge theory, the electron chemical potential $\mu_e$ 
and the color chemical potential $\mu_8$ are induced 
by the dynamics of gauge fields \cite{Gerhold2003}. 
The Maxwell and the Yang-Mills equations 
with the requirement of vanishing photon- and gluon-tadpole diagrams 
ensure electric and color neutrality. 
In NJL-type models, on the other hand, 
one has to impose the neutrality conditions 
by adjusting $\mu_e$ and $\mu_8$ \cite{BubSho}. 
In what follows, we neglect the color chemical potential 
because it is suppressed in the 2SC/g2SC phases, 
$\mu_8\sim{\cal O}(\Delta^2/\bar{\mu})\ll\Delta$.

In Nambu-Gor'kov space, the inverse full quark propagator 
$S^{-1}(p)$ is written as
\begin{eqnarray}
S^{-1}(p)=\left(
\begin{array}{cc}
(S_0^+)^{-1} & \Phi^- \\
\Phi^+ & (S_0^-)^{-1} 
\end{array}
\right),
\end{eqnarray}
with
\begin{subequations}
\label{eqn:CD}
\begin{eqnarray}
&&(S_0^+)^{-1}=\gamma^{\mu}p_{\mu}+(\bar{\mu}-\delta\mu\tau^3)\gamma^0
+g\gamma^{\mu}A_{\mu}^{a}T^{a},\\
&&(S_0^-)^{-1}=\gamma^{\mu}p_{\mu}-(\bar{\mu}-\delta\mu\tau^3)\gamma^0
-g\gamma^{\mu}A_{\mu}^{a}T^{aT},
\end{eqnarray}
\end{subequations}
and
\begin{eqnarray}
\Phi^- = -i\varepsilon\epsilon^b\gamma_5\Delta~,\qquad
\Phi^+ = -i\varepsilon\epsilon^b\gamma_5\Delta.
\end{eqnarray}
Here $\tau^3=\mbox{diag}(1,-1)$ is a matrix in flavor space. 
Following the usual convention, we choose the diquark condensate 
to point in the third direction in color space. 
In this work, we are interested in the Meissner masses squared of 
gluons of color 4--7 and 8, it is sufficient to study the case of 
the nonvanishing vector condensates 
$\langle gA_z^6 \rangle \neq 0$ 
or $\langle gA_z^8 \rangle \neq 0$ \cite{Gorbar2005b}. 

In the one-loop approximation, the effective potential 
of two-flavor quark matter (without electrons) is given by
\begin{eqnarray}
V(\Delta,\langle gA_z^{\alpha}\rangle,\delta\mu,\mu,T)
&=&\frac{\Delta^2}{4G_D}-\frac{T}{2}\sum_{n=-\infty}^{\infty}
\int^{\Lambda}\frac{d^3p}{(2\pi)^3}\nonumber\\
&&\times\ln\mbox{Det}S^{-1}(i\omega_n,\vec{p}),
\label{eqn:omega}
\end{eqnarray}
where $\omega_n=(2n+1)\pi T$ are the Matsubara frequencies 
and $\alpha\in(6,8)$. 
Note that the dependence on the vector condensates 
enters through the covariant derivative 
in the quark propagator (\ref{eqn:CD}).

The Meissner mass squared in the 2SC/g2SC phases can be calculated from
\begin{eqnarray}
m_{M,\alpha}^2=
\frac{\partial^2 V}{\partial \langle gA_z^{\alpha}\rangle^2}
\bigg{|}_{\langle gA_z^{\alpha}\rangle=0}.\label{eqn:mm}
\end{eqnarray}
However, in order to obtain meaningful results, 
we have to subtract ultraviolet divergences 
from Eq. (\ref{eqn:mm}) (see below).

\subsection{Meissner mass squared of gluons of color 4--7}
In this subsection we investigate the gluonic phase and 
the Meissner mass of the 6th gluon at nonzero temperature. 
Let us first note that, neglecting terms of 
order ${\cal O}(\bar{\mu}^2/\Lambda^2)$ 
and ${\cal O}(\Delta^2/\bar{\mu}^2)$, 
one can show that the potential curvature 
at zero temperature is \cite{Gorbar2005,KRS2006}
\begin{eqnarray}
m_{M,6}^2&=&\frac{\bar{\mu}^2}{6\pi^2}
\Bigg[1-\frac{2\delta\mu^2}{\Delta^2}\nonumber\\
&&\qquad+2\frac{\delta\mu\sqrt{\delta\mu^2-\Delta^2}}{\Delta^2}
\theta(\delta\mu-\Delta)\Bigg].\label{eqn:mm6}
\end{eqnarray}
The result is consistent with the Meissner mass squared for gluons 4--7 
derived within the hard-dense-loop (HDL) approximation \cite{Huang2004}. 
However, this is not the case for the Meissner mass squared 
calculated directly from Eq. (\ref{eqn:omega}). 
The potential curvature suffers 
from ultraviolet divergences $\propto\Lambda^2$ 
and therefore we have to subtract them.

At zero temperature the subtraction term is given by
\begin{eqnarray}
\frac{\partial^2 V(0,B,0,0,0)}
{\partial B^2}\bigg{|}_{B=0}=-\frac{\Lambda^2}{3\pi^2},\label{eqn:cz}
\end{eqnarray}
where $B\equiv\langle gA_z^6\rangle$. 
(In this paper, we shall refer to the phase 
where $B \neq 0$ as the gluonic phase.)

At nonzero temperature, on the other hand, 
the subtraction term should depend on temperature. 
Indeed one finds
\begin{eqnarray}
&&\frac{\partial^2 V(0,B,\delta\mu,\mu,T)}
{\partial B^2}\bigg{|}_{B=0}\nonumber\\
&&=\frac{\Lambda^2}{12\pi^2}\sum_{e1,e2,e3=\pm}
e_1N_F(e_1\Lambda+e_2\bar{\mu}+e_3\delta\mu),\label{eqn:ct}
\end{eqnarray}
where $N_F(x)=1/(e^{x/T}+1)$, 
otherwise the Meissner mass in the normal phase 
assumes an unphysical positive value. 
Let us note that, in the case of $T \to 0$, 
Eq. (\ref{eqn:ct}) is in exact agreement with Eq. (\ref{eqn:cz}). 
It should be also noted that the temperature dependence 
in Eq. (\ref{eqn:ct}) is made redundant 
when $\Lambda \gg T$. 
Thus, the temperature dependence of Eq. (\ref{eqn:ct}) 
is a cutoff artifact indeed. 

{\it In order to investigate the phase transition from 
the gluonic phase to the 2SC/g2SC phases}, 
we define the following normalized effective potential, 
that corresponds to the subtraction (\ref{eqn:ct}),
\begin{eqnarray}
\Omega_B\equiv V(\Delta,B,\delta\mu,\mu,T)
-V(0,B,\delta\mu,\mu,T).\label{eqn:omega2}
\end{eqnarray}
It should be noted that, even if $T \to 0$ or $\Lambda \gg T$, 
the effective potential (\ref{eqn:omega2}) 
itself is no longer coincident with 
that in our previous paper \cite{KRS2006}:
\begin{eqnarray}
\Omega_{B}' = V(\Delta,B,\delta\mu,\mu,T)
-V(0,B,0,0,0).\label{eqn:omega3}
\end{eqnarray}
If we introduce a cutoff large enough to neglect the temperature dependence 
in Eq. (\ref{eqn:ct}), we can use the mass subtraction (\ref{eqn:cz}) 
and, therefore, the potential subtraction (\ref{eqn:omega3}). 
(Of course, strictly speaking, such a cutoff should be infinite.) 
However, we wish to study the qualitative properties of 
the chromomagnetic instability in the phase diagram at morerate density regime 
within the gauged NJL model 
(i.e., nonrenormalizable four-fermi interactions), 
so we shall use the finite cutoff and, hence, 
the normalization (\ref{eqn:omega2}). 
Fortunately, we did not find a qualitative difference between 
$\Omega_B$ and $\Omega_{B}'$. 
Here we quote quantitative differences 
between these two normalizations at zero temperature: 
$B$ determined from Eq. (\ref{eqn:omega2}) around the gapless onset 
could be 40\% larger as compared to that in Ref. \cite{KRS2006} 
and, accordingly, the free energy gained by $B \neq 0$ could be 
35\% larger. 

\begin{figure}
\includegraphics[width=0.48\textwidth]{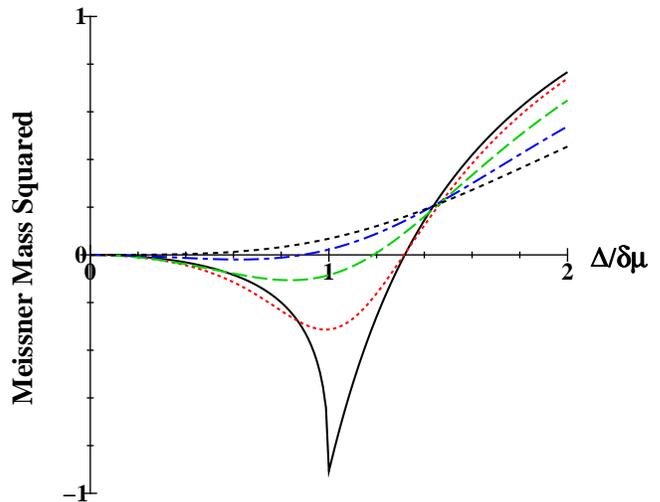}
\caption{The Meissner mass squared of the 6th gluon 
[divided by $\bar{\mu}^2/(6\pi^2)$] 
as a function of $\Delta/\delta\mu$ for $T=0$ MeV (solid), 
$T=10$ MeV (dotted) $T=20$ MeV (dashed), 
$T=30$ MeV (dot-dashed) and $T=40$ MeV (short-dashed). 
We used $\bar{\mu}=500$ MeV and $\delta\mu=80$ MeV.}
\label{Figure1}
\end{figure}
\begin{figure}
\includegraphics[width=0.45\textwidth]{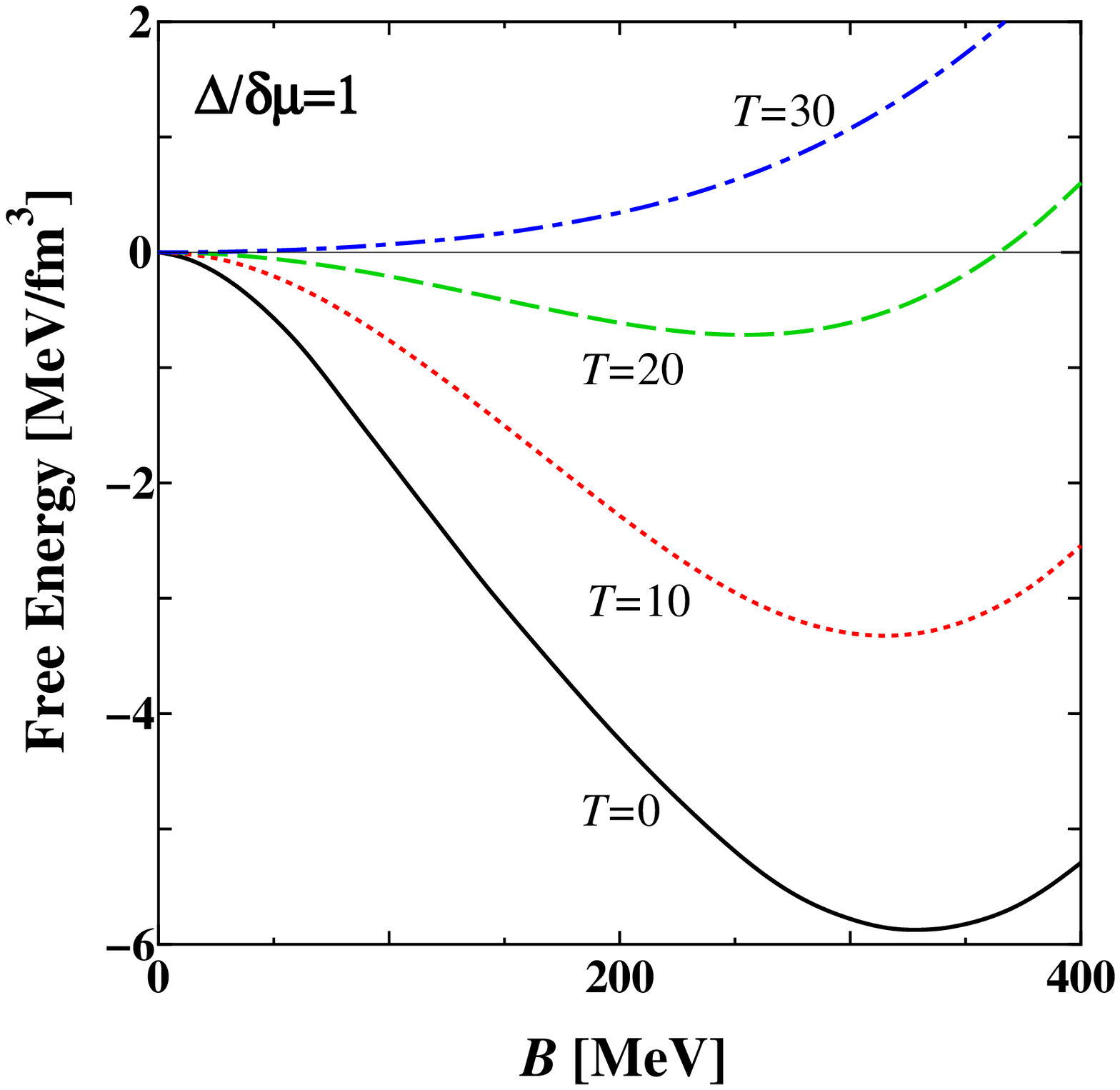}
\caption{The free energy measured with respect the 2SC/g2SC phases at $B=0$ 
as a function of $B$ for $T=0$ MeV (solid), 10 MeV (dotted), 
20 MeV (dashed) and 30 MeV (dot-dashed). 
We used the $T$-independent gap $\Delta=\delta\mu$. 
Other parameters are the same as in Fig. \ref{Figure1}.}
\label{Figure2}
\end{figure}
\begin{figure}
\includegraphics[width=0.45\textwidth]{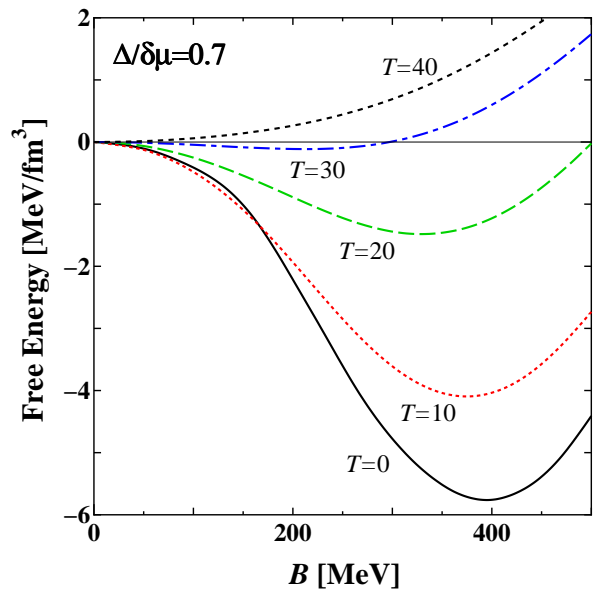}
\caption{The free energy measured with respect the 2SC/g2SC phases at $B=0$ 
as a function of $B$ for $T=0$ MeV (solid), 10 MeV (dotted), 
20 MeV (dashed), 30 MeV (dot-dashed) and 40 MeV (short-dashed). 
We used the $T$-independent gap $\Delta=0.7\delta\mu$. 
Other parameters are the same as in Fig. \ref{Figure1}.}
\label{Figure3}
\end{figure}

\begin{figure}
\includegraphics[width=0.45\textwidth]{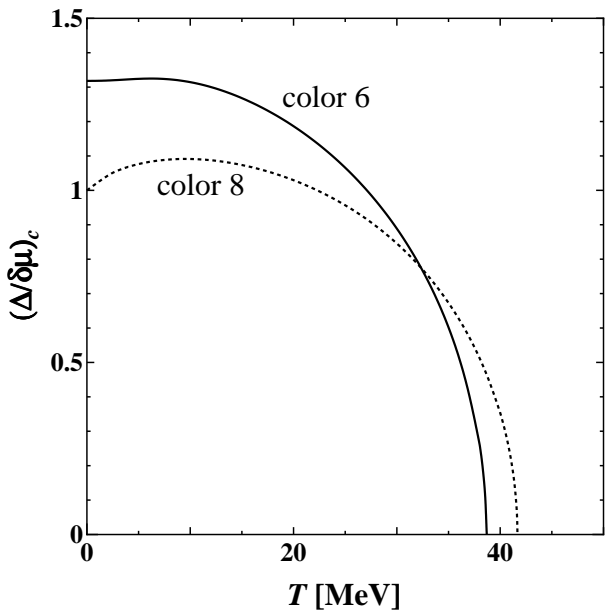}
\caption{The temperature dependence of the critical point 
$(\Delta/\delta\mu)_c$ of the chromomagnetic instability 
for gluons of color 4--7 (solid) and 8 (dotted). 
We used the same values of $\bar{\mu}$ 
and $\delta\mu$ as Fig. \ref{Figure1}.}
\label{Figure4}
\end{figure}

Figure \ref{Figure1} shows the Meissner mass squared of the 6th gluon, 
which is given by
\begin{eqnarray}
m_{M,6}^2=\frac{\partial^2\Omega_B(\Delta,B,\delta\mu,\mu,T)}
{\partial B^2}\bigg{|}_{B=0}.\label{eqn:MM6}
\end{eqnarray}
At $T=0$, we see the manifestation 
of the chromomagnetic instability at all values below 
$\Delta/\delta\mu=\sqrt{2}$. 
Note that the critical point of 
the instability $(\Delta/\delta\mu)_c$ 
is somewhat lower than $\sqrt{2}$. 
This is because our model parameters do not correspond to 
the HDL limit, so the contribution from subleading logarithms 
$\sim\Delta^2\ln(\Lambda/\Delta)$ is not negligibly small. 
We find, however, that the behavior of the Meissner mass squared 
is qualitatively consistent with that derived 
by using the HDL approximation \cite{Huang2004}.

As $T$ is increased, due to thermal smoothing effects, 
the characteristic kink at $\Delta/\delta\mu=1$ is smeared and 
the Meissner mass tends to approach its value in the normal phase. 
However, its temperature dependence 
(at fixed $\Delta/\delta\mu$) is non-monotonic. 
One can also see that the Meissner mass squared 
at small but nonzero values of $\Delta/\delta\mu$ 
remains negative even at high temperatures. 
At $T\simeq\delta\mu/2\simeq 40$ MeV, 
the chromomagnetic instability 
related to gluons 4--7 finally disappears 
and the Meissner mass squared turns positive 
at all values of $\Delta/\delta\mu$. 
As $T$ increases further, the Meissner mass squared begins 
to go down and approaches zero.

In Figs. \ref{Figure2} and \ref{Figure3}, 
we plot the effective potential (\ref{eqn:omega2}) 
measured with respect to the 2SC/g2SC phases at $B=0$ 
as a function of $B$ for several temperatures.

Figure \ref{Figure2} shows the case of $\Delta/\delta\mu=1$ 
(i.e., at the onset of the g2SC phase). 
At $T=0$, one clearly sees that the gluonic phase at $B \neq 0$ 
is energetically more favored than the g2SC phase at $B=0$. 
As $T$ is increased, the vacuum expectation value (VEV) of $B$ 
continuously goes to zero and the free energy gained 
by the gluonic phase also decreases to zero. 
We observe that the second-order phase transition 
from the gluonic phase to the g2SC phase 
occurs at $T \simeq 30$ MeV. 
It can be also seen that the potential curvature at $B=0$ 
is negative at $T=0$, 
corresponding to the negative Meissner mass squared 
in the g2SC phase at $T=0$. 
The potential curvature grows with increasing temperature 
(within the interval $T=0 \to 30$ MeV) 
and its temperature dependence agrees 
with the result shown in Fig. \ref{Figure1}. 

In Fig. \ref{Figure3}, the same plot is shown for 
the case of $\Delta/\delta\mu=0.7$. 
At this value of $\Delta/\delta\mu$, 
the temperature dependence of the Meissner mass squared 
is not monotonic (see Fig. \ref{Figure1}). 
One can in fact see that, as temperature grows, the potential curvature 
at $B=0$ first drops and then goes up. 
On the other hand, the free energy gain monotonically 
decreases with increasing temperature. 
We again find that the second-order phase transition 
at $T \simeq 35$ MeV. 

Here, we would like to make some comments. 
The phase transition from the gluonic to the 2SC/g2SC phases 
is, presumably, of second order. 
(Evaluating the effective potential 
at small $\Delta/\delta\mu$'s with sufficient accuracy is not easy 
and hence we cannot exclude the possibility of a first-order transition.) 
If so, the Meissner mass squared in the 2SC/g2SC phase 
could be a criterion for choosing the energetically favored phase 
without calculating the free energy. 

In Fig. \ref{Figure4}, we plot the temperature dependence 
of the critical point $(\Delta/\delta\mu)_c$ 
where the Meissner mass becomes negative. 
It is seen that $(\Delta/\delta\mu)_c$ continuously goes to zero 
at $T \simeq \delta\mu/2$. 
We have checked the robustness of this relation 
varying model parameters and found that it works well 
as long as $\bar{\mu}$ is not too small. 
One can also see that $(\Delta/\delta\mu)_c$ 
grows slightly at low temperatures. 
Using huge values of $\Lambda$ and $\mu$, 
we confirmed that this behavior remains true even in the HDL limit.

\subsection{Meissner mass squared of gluons of color 8}
\begin{figure}
\includegraphics[width=0.48\textwidth]{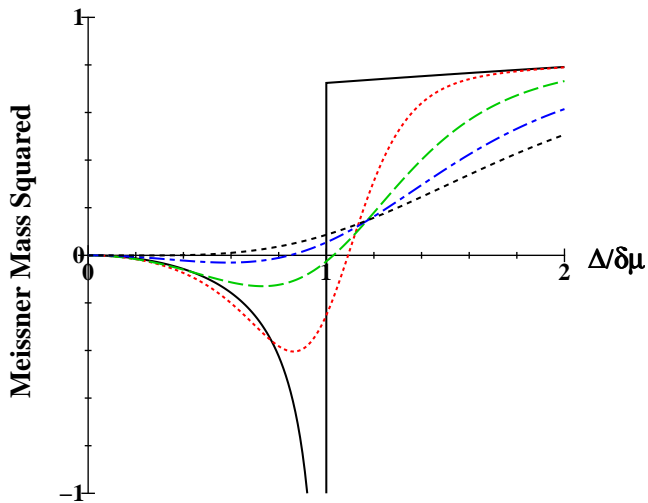}
\caption{The Meissner mass squared 
of the 8th gluon [divided by $\bar{\mu}^2/(6\pi^2)$] 
as a function of $\Delta/\delta\mu$ for 
$T=0$ MeV (solid), $T=10$ MeV (dotted) $T=20$ MeV (dashed), 
$T=30$ MeV (dot-dashed) and $T=40$ MeV (short-dashed). 
We used the same values of $\bar{\mu}$ 
and $\delta\mu$ as in Fig. \ref{Figure1}.}
\label{Figure5}
\end{figure}
\begin{figure}
\includegraphics[width=0.45\textwidth]{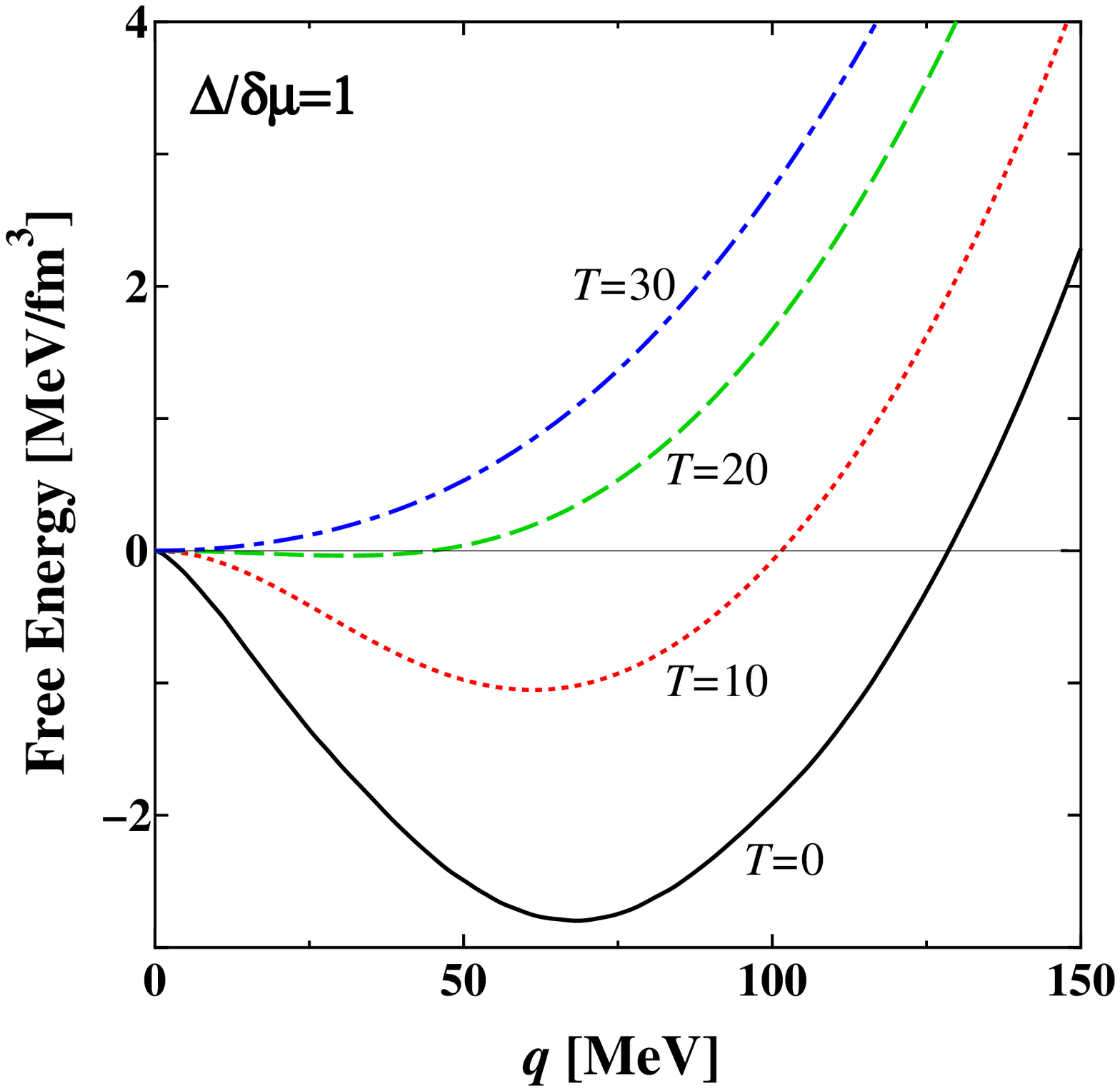}
\caption{The free energy measured with respect the 2SC/g2SC phases at $q=0$ 
as a function of $q$ for $T=0$ MeV (solid), 10 MeV (dotted), 
20 MeV (dashed) and 30 MeV (dot-dashed). 
We used the $T$-independent gap $\Delta=\delta\mu$ 
and the same values of $\bar{\mu}$ and $\delta\mu$ as in Fig. \ref{Figure1}.}
\label{Figure6}
\end{figure}
\begin{figure}
\includegraphics[width=0.45\textwidth]{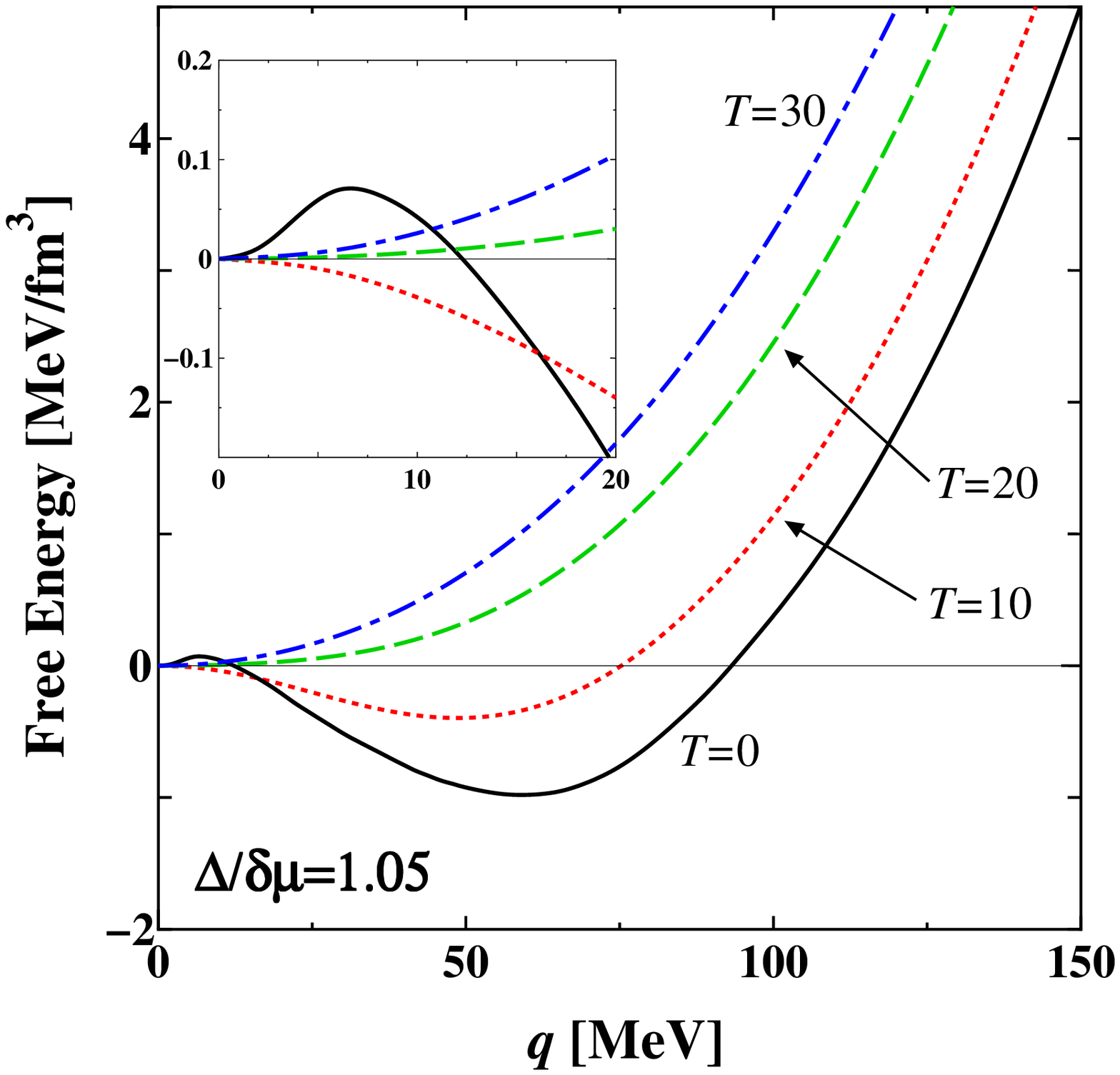}
\caption{The free energy measured with respect the 2SC/g2SC phases at $q=0$ 
as a function of $q$ for $T=0$ MeV (solid), 10 MeV (dotted), 20 MeV (dashed), 
and 30 MeV (dot-dashed). 
We used the $T$-independent gap $\Delta=1.05\delta\mu$ 
and the same values of $\bar{\mu}$ and $\delta\mu$ as in Fig. 1.}
\label{Figure7}
\end{figure}
We now turn to the 8th gluon. 
In order to investigate the single plane-wave LOFF state, 
we use the following gauge transformation 
described in Ref. \cite{Gorbar2005b}. 
Using the gauge transformation 
$\psi\to\psi'=\exp(-i\vec{q}\cdot\vec{x})\psi$, 
one can show that the single plane-wave LOFF state 
whose gap parameter spatially oscillates like 
$\Delta(x)=\Delta\exp(2i\vec{q}\cdot\vec{x})$ 
is equivalent to the 2SC/g2SC phases 
with the Abelian vector condensate 
$\vec{q}=\langle g\vec{A}^8\rangle/(2\sqrt{3})$. 

The same argument for the normalization of the effective potential 
that we made in the previous subsection 
holds also for the 8th gluon. 
Therefore the normalized effective potential 
for the single plane-wave LOFF state 
at nonzero temperature is
\begin{eqnarray}
\Omega_q \equiv V(\Delta,q,\delta\mu,\mu,T)
-V(0,q,\delta\mu,\mu,T).\label{eqn:omegaq}
\end{eqnarray}
where $q\equiv\langle gA_z^8\rangle/(2\sqrt{3})$. 
Here, we chose the third component of $\vec{q}$ 
without loss of generality. 
The Meissner mass squared in the 2SC/g2SC phases is then given by
\begin{eqnarray}
m_{M,8}^2=\frac{1}{12}\frac{\partial^2\Omega_q(\Delta,q,\delta\mu,\mu,T)}
{\partial q^2}
\bigg{|}_{q=0}.\label{eqn:MM8}
\end{eqnarray}
We have checked that the subtraction term,
\begin{eqnarray}
\frac{1}{12}\frac{\partial^2 V(0,q,\delta\mu,\mu,T)}{\partial q^2}\bigg{|}_{q=0},
\end{eqnarray}
agrees with $-\Lambda^2/(9\pi^2)$ when $T \to 0$ or $\Lambda \gg T$.

In Fig. \ref{Figure5}, we plot $m_{M,8}^2$ 
as a function of $\Delta/\delta\mu$ 
for several temperatures. We see that $m_{M,8}^2$ at $T=0$ shows 
different behavior from 
that obtained by using the HDL approximation \cite{Huang2004}: 
in particular, in the 2SC phase, $m_{M,8}^2$ is not constant and 
there exists an enhancement in its values 
(the value of $m_{M,8}^2$ in the HDL approximation 
should be 2/3 in Fig. \ref{Figure5}). 
This is because, in Fig. \ref{Figure5}, 
contributions from subleading logarithms 
have not been subtracted from the potential curvature. 
However, the behavior is qualitatively consistent with the HDL result 
and, furthermore, we observe a negative infinite Meissner mass squared 
at the gapless onset $\Delta/\delta\mu=1$. 

At $T>0$, the negative infinite Meissner mass squared at the gapless onset is 
smeared and $m_{M,8}^2$ tends to approach zero. 
However, the temperature dependence is non-monotonic. 
We also find a temperature-induced instability 
in a narrow region above $\Delta/\delta\mu=1$. 
(In Fig. \ref{Figure4}, one can see that 
the critical point for the 8th gluon grows at low temperatures. 
We shall take a closer look at this problem later.) 
These same results has been observed in Ref. \cite{Alford2005}. 
Like in the case for gluons 4--7, 
$m_{M,8}^2$ at small $\Delta/\delta\mu$'s remains negative 
at high temperatures, 
but the instability related to the 8th gluon 
disappears at $T \simeq \delta\mu/2$.

In Figs. \ref{Figure6} and \ref{Figure7}, 
we illustrate the effective potential measured with respect to 
the 2SC/g2SC phases at $q=0$ 
as a function of $q$ for several temperatures. 

Figure \ref{Figure6} corresponds to the gapless onset $\Delta/\delta\mu=1$. 
At $T=0$, the LOFF state at $q \simeq 70$ MeV
is energetically more favored state than the g2SC phase
and the potential curvature at $q=0$ has a cusp,
corresponding to a negative infinite Meissner mass squared
at $\Delta/\delta\mu=1$.
(The result is consistent with that reported in Ref. \cite{Fukush2006}.) 
The cusp at $q=0$ is immediately smeared by temperature effects
and the Meissner mass squared takes negative nonzero values.
As $T$ is increased, both the VEV of $q$ and the free-energy gain
by the LOFF state are decreased. 
At $T \simeq 30$ MeV, we find the second-order phase transition from
the LOFF state to the g2SC phase. The temperature dependence
of the potential curvature at $q=0$ is
consistent with the results shown in Fig. \ref{Figure5}.

Figure \ref{Figure7} shows the temperature dependence of
the effective potential
at $\Delta/\delta\mu=1.05$, slightly above the gapless onset. 
In this case, as mentioned earlier,
one finds the instability only at $T>0$. 
In terms of the free energy,
the reason for the temperature-induced instability
can be interpreted as follows. 
At $\Delta/\delta\mu=1.05$, the Meissner mass squared in the 2SC phase
is positive at $T=0$ (see Fig. \ref{Figure5}). 
One can see that the potential curvature at $q=0$ is indeed
positive at $T=0$. 
However, as is clear from Fig. \ref{Figure7},
there exists an energetically more favored state at $q \neq 0$,
the LOFF state. 
Therefore, the 2SC phase is only metastable,
though the Meissner mass squared is positive in this phase. 
(The metastable 2SC phase,
separated from the LOFF state by a potential hump,
exists in the region $1 < \Delta/\delta\mu \alt 1.08$.) 
As $T$ grows, the potential hump is smoothed out and,
at a certain small temperature, the 2SC phase becomes unstable against
the formation of the LOFF state. 
Specifically the Meissner mass squared turns negative at $T \simeq 2$ MeV.
At $T \simeq 19$ MeV, we find a second-order phase transition
from the LOFF state to the 2SC phase.
The temperature dependence of the effective potential
clearly explains the induced instability.
However, it is fair to say that
we have found an unexpected growth of the critical point
also for the 8th gluon: $(\Delta/\delta\mu)_c$ grows to
$\Delta/\delta\mu \simeq 1.09$.

Before concluding this subsection, 
we would like to mention 
the order of the phase transition LOFF $\to$ 2SC/g2SC. 
We found a second-order phase transition 
in a wide range of $\Delta/\delta\mu$. 
However, it is not easy to determine 
the order of the transition at small $\Delta/\delta\mu$ 
and around $\Delta/\delta\mu \simeq 1.08$. 
Hence, we do not exclude the possibility of a first-order transition.

\subsection{Phase diagram in $T$-$\Delta_0$ plane}
\begin{figure*}
\includegraphics[width=0.98\textwidth]{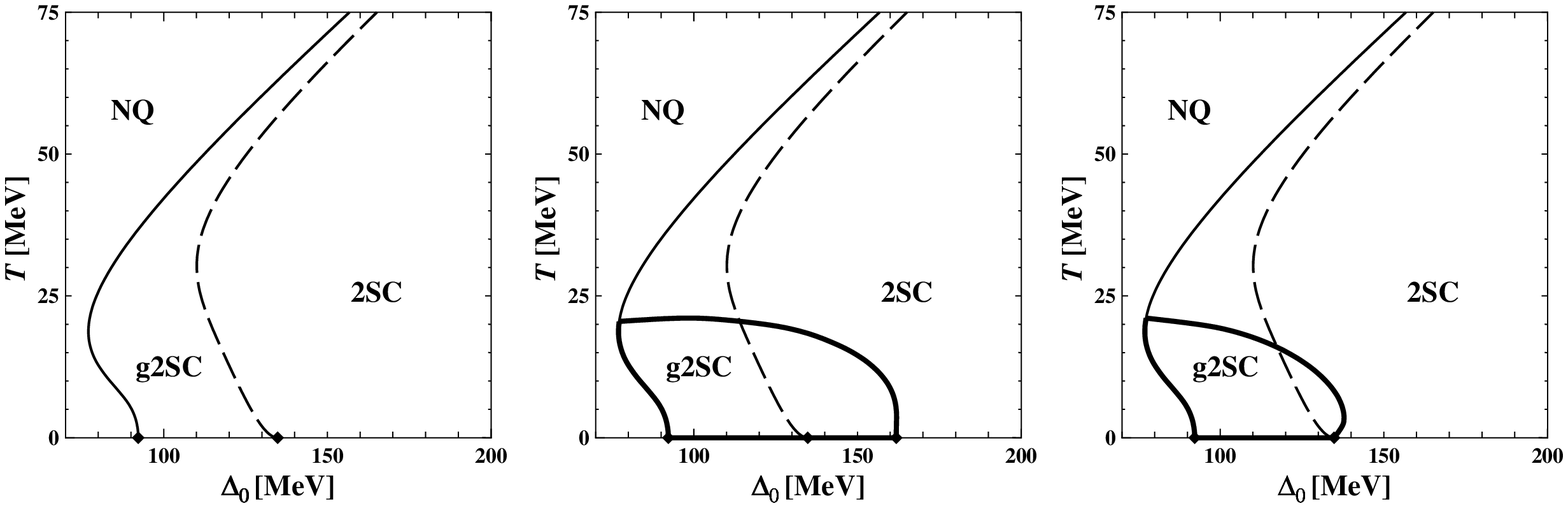}
\caption{Left: The phase diagram of a neutral two-flavor color superconductor 
in the plane of temperature and $\Delta_0$. 
The solid (dashed) line denotes the critical line of 
the phase transition between the normal quark phase and the g2SC phase 
(the g2SC phase and the 2SC phase). 
The results are plotted for $\mu=400$ MeV. 
Middle: The same as the left panel, but the unstable region for 
gluons 4--7 is depicted by the region enclosed by the thick solid line. 
Right: The same as the left panel, but the unstable region for 
the 8th gluon is depicted by the region enclosed by the thick solid line.}
\label{Figure8}
\end{figure*}
\begin{figure}
\includegraphics[width=0.44\textwidth]{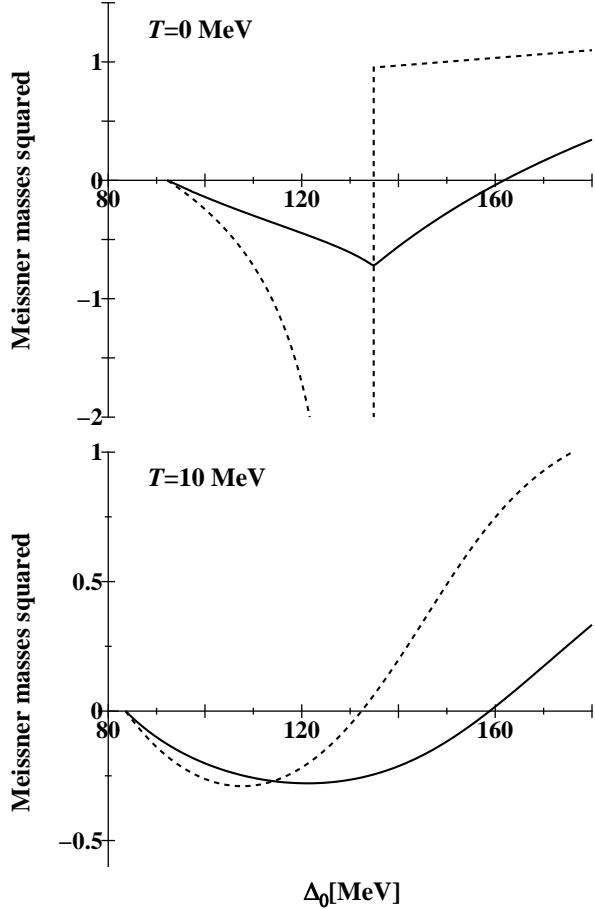}
\caption{The Meissner masses squared of gluons of color 6 (solid) 
and 8 (dotted) [divided by $\bar{\mu}^2/(6\pi^2)$] as a function of 
$\Delta_0$ for $T=0$ MeV (upper panel) and $T=10$ MeV (lower panel). 
The results are plotted for $\mu=400$ MeV.}
\label{Figure9}
\end{figure}
\begin{figure}
\includegraphics[width=0.44\textwidth]{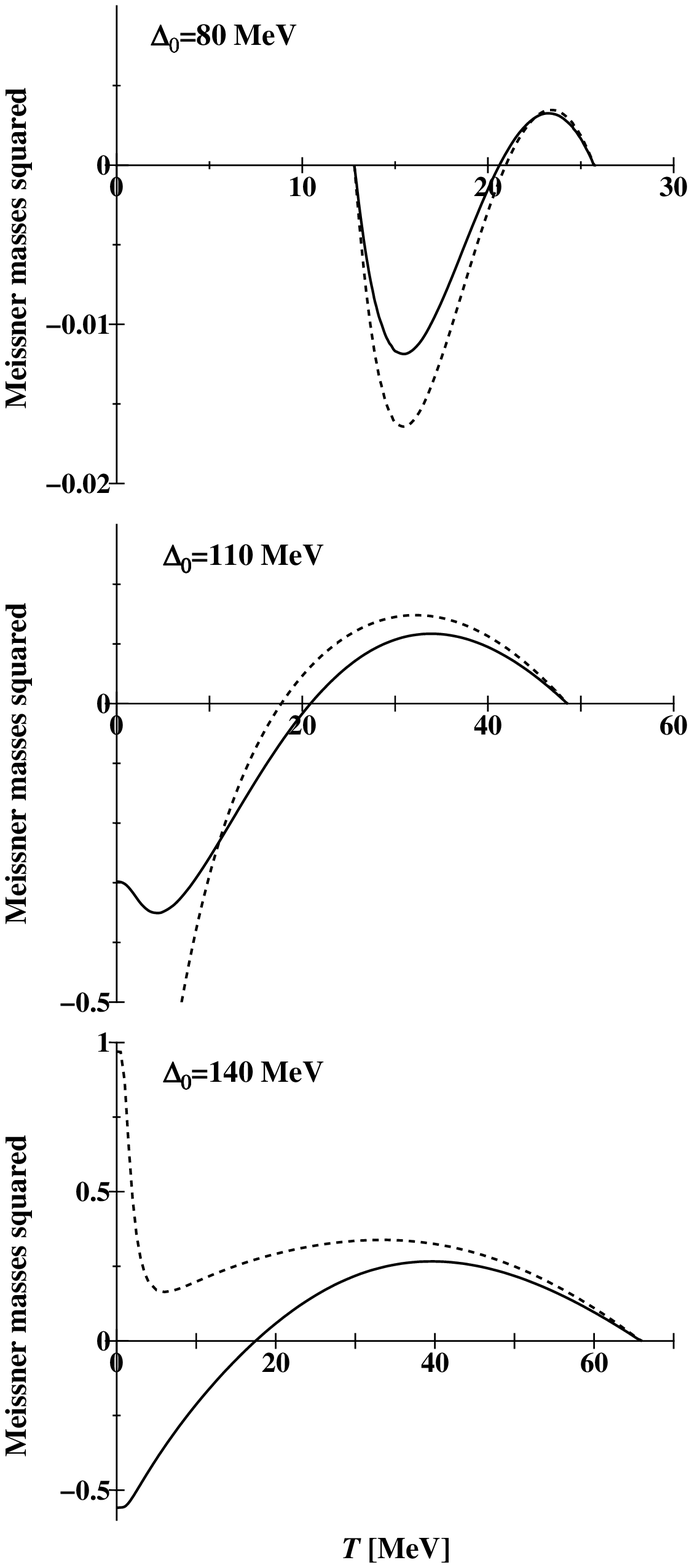}
\caption{The Meissner masses squared of gluons of color 6 (solid) 
and 8 (dotted) [divided by $\bar{\mu}^2/(6\pi^2)$] as a function of $T$ for 
$\Delta_0=80$ MeV (upper panel), $\Delta_0=110$ MeV (middle panel) and 
$\Delta_0=140$ MeV (lower panel). 
The results are plotted for $\mu=400$ MeV.}
\label{Figure10}
\end{figure}
In order to see the consequences of the present analysis, 
we study the phase diagram of a two-flavor color superconductor. 
To this end, we first solve the gap equation,
\begin{eqnarray}
\frac{\partial\Omega_{\mb{2SC/g2SC}}}{\partial\Delta}=0,\label{eqn:gap}
\end{eqnarray}
and the neutrality condition,
\begin{eqnarray}
\frac{\partial\Omega_{\mb{2SC/g2SC}}}{\partial\mu_e}=0,\label{eqn:enc}
\end{eqnarray}
where the effective potential of the neutral 2SC/g2SC phases 
$\Omega_{\mb{2SC/g2SC}}$ is given by
\begin{eqnarray}
\Omega_{\mb{2SC/g2SC}}&=&-\frac{1}{12\pi^2}
\left(\mu_e^4+2\pi^2T^2\mu_e^2+\frac{7\pi^4}{15}T^4\right)\nonumber\\
&&+V(\Delta,0,\delta\mu,\mu,T).\label{eqn:omega4}
\end{eqnarray}
Here, the contribution from electrons has been added to Eq. (\ref{eqn:omega}). 

In the left panel of Fig. \ref{Figure8}, 
we illustrate the phase diagram of a neutral two-flavor color superconductor 
in the plane of $T$ and $\Delta_0$, 
where $\Delta_0$ is the value of the 2SC gap at $\delta\mu=0$ and at $T=0$ 
(i.e., the parameter $\Delta_0$ corresponds 
to the diquark coupling strength). 
The result is plotted for $\mu=400$ MeV, which is 
a value typical for the interior of compact stars. 
The solid (dashed) line denotes the critical line of 
the phase transition between the normal quark phase and the g2SC phase 
(the g2SC phase and the 2SC phase). 

Let us have a look at main features of the phase diagram. 
One sees that qualitative features of the left panel of Fig. \ref{Figure8} 
are consistent with the $T$-$\mu$ phase diagram in the literature.

In the weak-coupling regime $77~{\rm MeV}<\Delta_0<92~{\rm MeV}$, 
the Fermi momentum mismatch is too large 
for these coupling strengths for diquark pairing 
and the system is in the normal quark (NQ) phase at $T=0$. 
At $T>0$, the mismatch of the Fermi surfaces is thermally smeared 
and, then, it opens the possibility of finding the g2SC phase 
(see, for example, Fig. 1 in the first paper in Ref. \cite{phased2} 
and Fig.4 in the third paper in Ref. \cite{phased3}).

In the intermediate coupling regime, $92~{\rm MeV}<\Delta_0<134~{\rm MeV}$, 
the g2SC phase is realized at $T=0$. 
For relatively strong coupling, $110~{\rm MeV}<\Delta_0<134~{\rm MeV}$, 
the g2SC phase at low temperature is replaced by the 2SC phase 
at intermediate temperature. 
At higher temperature, the 2SC phase is replaced by the g2SC phase again. 
It is known that this unusual behavior happens 
in the intermediately coupled two-flavor quark matter. 
For a detailed discussion, see the second paper in Ref. \cite{Shovkovy2003}.

For strong coupling, $\Delta_0>134~{\rm MeV}$, 
the gap $\Delta$ increases and the 2SC phase is 
accordingly favored at $T=0$. 
At higher temperatures, however, $\Delta$ is decreased by thermal effects 
and the g2SC phase becomes possible 
(cf. Fig. 6 in the third paper in Ref. \cite{phased2}).

Let us now take into account the chromomagnetic instability. 
Combining Eqs. (\ref{eqn:gap}) and (\ref{eqn:enc}) 
with Eqs. (\ref{eqn:MM6}) and (\ref{eqn:MM8}), 
we calculate the Meissner masses squared 
and map out the unstable regions on the phase diagram. 
(It should be remembered that we do not solve the gap equations 
for $\Delta$ and $\langle gA_z^{\alpha} \rangle$ 
and the neutrality condition for $\mu_e$ self-consistently.) 

The region enclosed with the solid thick line 
in the middle panel of Fig. \ref{Figure8} 
corresponds to the unstable region 
where gluons 4--7 have a negative Meissner mass squared. 
In this unstable region, therefore, the gluonic phase should be realized 
by a nonvanishing VEV of $\langle gA_z^6 \rangle$. 
At $T=0$, we see the manifestation of the instability 
in the region $92~{\rm MeV}<\Delta_0<162~{\rm MeV}$ 
(see the upper panel of Fig. \ref{Figure9}). 
(Note that the upper boundary of this unstable region, 
$\Delta_0=162~{\rm MeV}$, is 
lower than that quoted in Refs. \cite{Gorbar2005b,KRS2006}. 
The reason for this discrepancy is the following: 
in order to find the unstable window, we have calculated 
the Meissner mass squared directly, 
whereas they used the relation $(\Delta/\delta\mu)_c=\sqrt{2}$, 
which is derived by using the HDL approximation. 
Of course, the discrepancy is nothing but a cutoff artifact.) 
At low temperatures, the whole g2SC phase 
and a part of the 2SC phase suffer from the instability. 
At $T \simeq 20$ MeV, the instability 
related to gluons 4--7 is washed out and 
the phase transition is most probably of second order. 
(The unstable region should disappear at $T \simeq \delta\mu/2$ 
along the thin solid line (see Fig. \ref{Figure4}). 
As a check, let us assume $\mu_e \simeq 90$ MeV, 
which roughly corresponds to the value of $\mu_e$ 
at the zero-temperature edge of the g2SC window. 
Then we see that the relation yields $T \simeq 20$ MeV indeed.)

In the right panel of Fig. \ref{Figure8}, 
the unstable region for the 8th gluon 
is depicted by the enclosed region. 
(The region in which the 2SC phase is metastable 
is not shown in this figure.) 
In this region, the LOFF state is favorable 
to cure the instability related to the 8th gluon. 

At $T=0$, as it should be, only the g2SC phase suffers from the instability. 
At $T>0$, however, the critical point shifts to 
larger $\Delta_0$'s and the unstable region penetrates into the 2SC phase. 
(This is nothing but the temperature-induced instability.) 
While the temperature-induced instability 
in the 2SC phase disappears at $T \simeq 16$ MeV, 
the unstable region in the g2SC phase remains until $T \simeq 21$ MeV.

Overlapping the middle and the right panels 
of Fig. \ref{Figure8}, 
one can see that, apart from the case of strong coupling, 
the 2SC/g2SC phases at low temperature are unstable. 
In particular, the g2SC phase suffers from a severe instability 
related to both gluons 4--7 and 8. 
(In Fig. \ref{Figure10}, we illustrate 
the temperature dependence of the Meissner masses squared 
$m_{M,6}^2$ and $m_{M,8}^2$ for the cases of $\Delta_0=80,~110,~140$ MeV. 
It is clear that at least one of the the Meissner masses 
in the 2SC/g2SC phases is imaginary 
at low temperatures and both Meissner masses squared 
become positive at $T \simeq 20$ MeV.) 
For strong couplings $\Delta_0 \agt 162~{\rm MeV}$, 
the system gets rid of the chromomagnetic instability. 
Note that one still finds a stable g2SC phase 
in the high-temperature region in the phase diagram, 
though the gapless structure 
is not significant at high temperature.

\section{Summary and discussion}
We studied the chromomagnetic instability 
in two-flavor quark matter at nonzero temperature. 
We use the gauged NJL model and first analyzed 
the curvature of the effective potential. 
Then, a temperature-dependent subtraction for the Meissner masses 
squared was introduced, so that magnetic gluons remain unscreened 
in the normal phase. As mentioned earlier, 
this temperature dependence is indeed the cutoff artifact. 
In this work, we studied the properties of 
the chromomagnetic instability 
in the intermediate and strong coupling regimes, 
using a phenomenological four-fermi interaction. 
Therefore, we used a temperature-dependent subtraction 
and, accordingly, the ad hoc normalization of the effective potential. 
It should be emphasized that, 
for our standard value of the NJL cutoff, 
the temperature dependence of the subtraction term 
is not negligibly small actually. 

We calculated the temperature dependence of the Meissner masses 
squared as a function of $\Delta/\delta\mu$ 
and found that, at $T \simeq \delta\mu/2$, the instability related 
to gluons 4--7 and 8 is washed out at all values of $\Delta/\delta\mu$. 
We also confirmed the temperature-induced instability 
(i.e., the growth of the critical point at $T>0$) for the 8th gluon. 
In order to look at the temperature dependence of the 
Meissner masses squared, we computed 
not only the potential curvature but also 
the effective potential itself as a function of the vector 
condensates $\langle gA_z^6 \rangle$ and $\langle gA_z^8 \rangle$. 
Evaluating the effective potential played a crucial role 
for understanding the temperature-induced instability for the 8th gluon. 
By comparing the free energies of the 2SC phase 
and the single plane-wave LOFF state, 
we clarified that the induced instability mainly arises 
from the fact that the 2SC phase 
in the region slightly above the gapless onset 
is only metastable at $T=0$. 

We also presented the phase diagram in the $T$-$\Delta_0$ plane 
and mapped out the unstable regions 
for gluons 4--7 and 8 on the phase diagram. 
We found that, apart from the case of strong coupling, 
the 2SC/g2SC phases at low temperatures $\alt 20~{\rm MeV}$ 
suffer from a severe instability 
related to both gluons 4--7 and 8 and a large region 
in the g2SC phase should be replaced by the vector condensed phases. 

In calculating the effective potential, 
we did not enforce the neutrality condition. 
Hence, the effective potential shown 
in Figs. \ref{Figure2}, \ref{Figure3}, \ref{Figure6} 
and \ref{Figure7} might be altered by the neutrality constraint. 
The result for the LOFF state (Fig. \ref{Figure7}) 
should be taken as an indication that the neutral 2SC phase 
is not stable against the formation of the LOFF state even before 
$\Delta/\delta\mu$ reaches the gapless onset 
and that a first-order transition (2SC $\leftrightarrow$ LOFF) 
occurs at a certain value of $\Delta/\delta\mu>1$. 
However, they must remain true 
even if we take into account the neutrality condition. 
At $T=0$, a neutral LOFF state was studied 
by solving the gap equations for $\Delta$ and $q$ 
and the neutrality condition for $\mu_e$ self-consistently 
and it was revealed that the LOFF state is indeed 
favored over the 2SC phase even above the gapless onset 
(the edge of the LOFF state with the 2SC phase was 
determined to be $\Delta_0=137~{\rm MeV}$ \cite{Gorbar2005b}). 
In addition, Fig. 1 in Ref. \cite{Gorbar2005b} indicates 
such a first-order transition actually happens. 
At $T>0$, although we did not examine the low-temperature effective potential 
of the LOFF state, it is likely that 
the first-order transition (2SC $\leftrightarrow$ LOFF above the gapless onset) 
takes place as long as $T$ is not too high.

Let us look at the right panel of Fig. \ref{Figure8} again. 
(We should recall that the region where the 2SC is metastable 
is not depicted in Fig. \ref{Figure8}.) 
As we mentioned above, there exists a window 
where the LOFF state is energetically more favored than the 2SC phase, 
even though $m_{M,8}^2>0$ in the 2SC phase. 
The actual phase boundary (2SC $\leftrightarrow$ LOFF), 
as a consequence, shifts to larger $\Delta_0$'s 
when we solve the set of equations self-consistently 
and take account of the metastability of the 2SC phase. 
In addition, the NQ phase in the weak coupling regime 
will be replaced by the neutral LOFF state. 
At $T=0$, in fact, it was found that the neutral LOFF state exists 
in the window $63~{\rm MeV}<\Delta_0<137~{\rm MeV}$ 
and that it is more stable than the NQ phase 
in whole this window \cite{Gorbar2005b}. 
The weakly coupled LOFF state survives at nonzero temperature 
and, presumably, undergoes a phase transition into the g2SC phase 
or the NQ phase \cite{HJZ}. 
Then, we conclude that, apart from the case of strong coupling, 
the low-temperature ($\alt 20~{\rm MeV}$) region 
of a chromomagnetically stable, non color-flavor-locked phase 
has a completely different structure from known phase diagrams. 

The most interesting remaining task is now clear: 
we have to take into account all the possible gluonic condensates 
and calculate the free energy in a self-consistent manner. 
(In particular, we have a limited knowledge 
of the gluonic phase \cite{Gorbar2005,KRS2006}. 
While suggestive, it is not sufficient for drawing a conclusion 
about the phase structure of the gluonic phase.) 
The resulting gluonic-condensed phase will be free 
from the chromomagnetic instability. 
Here, it should be noted that the study of the effective potential 
shows that the Meissner masses squared itself 
cannot be a criteria for choosing the ground state, in other words, 
the small-$\langle A_z^{\alpha} \rangle$ 
expansion of the effective potential does not work. 
It is also interesting to make a (free energy) comparison 
between the gluonic-condensed phase and 
the LOFF state with realistic crystal structures. 
Finally, the gluonic-condensed phase has not been studied 
in the three-flavor case, 
so an extension to the realistic three-flavor quark matter 
would have important implications for the physics of compact stars.

\begin{acknowledgments}
The author would like to thank Igor Shovkovy for 
fruitful discussion and Dirk Rischke for comment 
on the earlier version of the manuscript. 
This work was supported 
by the Deutsche Forschungsgemeinschaft (DFG).

{\it Note added.} While writing this paper, 
I learned that an overlapping study was recently done by 
L. He, M. Jin, and P. Zhuang \cite{HJZ}. 
I am grateful that they made the results of their study available to me.
\end{acknowledgments}

\end{document}